\documentclass[a4paper]{jpconf}
\usepackage{graphicx}
\begin{document}
\title{Measurement of the $\eta$ and $\omega$ Dalitz decays transition form factors in p-A collisions at 400~GeV/$c$ with the NA60 apparatus}

\author{Antonio Uras for the NA60 Collaboration}

\address{University of Cagliari and INFN, Cagliari. Complesso Universitario di Monserrato - S.P.~per Sestu Km 0.700 09042 Monserrato (Cagliari) - Italy}

\ead{antonio.uras@ca.infn.it}

\begin{abstract}
The NA60 experiment has studied low-mass muon pairs production in proton-nucleus collisions at 400 GeV/$c$ at 
the CERN SPS. The mass spectrum is well described by the superposition of the two-body and Dalitz decays of the light neutral mesons $\eta$, 
$\rho$, $\omega$, $\eta'$ and $\phi$. The large collected statistics allows to isolate the contributions due to the Dalitz decays of the 
$\eta$ and $\omega$ mesons, from which the electromagnetic transition form factors of the two mesons have been extracted. The found values 
agree with the previous available measurements, improving their uncertainty thanks to the higher statistics. The results thus confirm the 
discrepancy with the prediction of the Vector Meson Dominance model in the case of the electromagnetic form factor of the $\omega$ meson.
\end{abstract}


\noindent Proton-nucleus studies have been part of the heavy-ion experimental program at the CERN SPS, of which the NA60 experiment
represents the most recent and one of the most fruitful efforts. Beyond serving as a robust reference baseline for the heavy-ion
data, proton-nucleus data also stimulate a genuine physics interest, allowing measurements in an environment of cold nuclear matter.
This is in fact the case of the electromagnetic transition form factors of the $\eta$ and $\omega$ mesons, which have been studied here through 
their Dalitz decays $\eta \to \mu^+ \mu^- \gamma$ and $\omega \to \mu^+ \mu^- \pi^0$. In these $A \to B l^+ l^-$ processes, the mesons decay electromagnetically
into a virtual photon with mass $M$ $-$ in turn converting into a lepton pair $-$ and a third body. Assuming point-like particles,
the mass distribution of the emitted lepton pair (that is, the decay rate of the process as a function of $M$) can be exactly 
described by QED as firstly shown by N.M.~Kroll and W.~Wada~\cite{KW} in the 1950s. 

However, the observed mass distribution for the lepton pairs emitted in an $A \to B l^+ l^-$ decay differs from the point-like one, because of
the internal electromagnetic structure of the decaying meson, which reveals at the vertex of the transition $A \to B$. This difference
can be formally described by a function of $M$, usually expressed as $|F_{AB}(M)|^2$, known as \emph{transition form factor}. Experimentally,
$|F_{AB}(M)|^2$ is directly accessible by comparing the measured invariant mass spectrum of the lepton pairs from the Dalitz decays,
with the point-like QED prediction. The transition form factors are usually given in terms of the pole parametrization:
\begin{equation}
 |F|^2 = (1 - M^2/\Lambda^2)^{-2}~.
\end{equation}
The Vector Meson Dominance (VMD) model gives theoretical predictions both for $\Lambda_{\eta}^{-2}$ and $\Lambda_{\omega}^{-2}$. These predictions have been tested by the Lepton-G experiment \cite{landsberg} using pion beams, and recently by the NA60 experiment in In-In peripheral collisions \cite{na60}. While the experimental values agree among each other both for the $\eta$ and $\omega$ mesons, only the VMD model prediction for the $\eta$ meson is confirmed by the data, while the one for the $\omega$ meson is not, strongly underestimating the observed form factor. The NA60 p-A data can now give a further significant contribution to the measurement, thanks to the high quality and the large statistics available.

~\\
\noindent A description of the NA60 apparatus can be found for example in \cite{apparatus}. Here, only few relevant details are given. The produced dimuons are identified and measured by the muon spectrometer, composed of a set of tracking stations, trigger scintillator hodoscopes, a toroidal magnet and a hadron absorber. The stopping of hadrons (provided by more than 20 interactions lengths) allows to select the highly rare dimuon events. On the other hand, the material which stops the hadrons also induces multiple scattering and energy loss on the muons, degrading the mass resolution of the measurement made in the spectrometer. To overcome this problem, NA60 already measures the muons before the absorber with a tracking telescope, made of pixel silicon detectors \cite{pixel}. The muon tracks are found among the many other charged particle tracks by a correct matching, both in angular and momentum coordinates, with the reconstructed tracks in the muon spectrometer.
%
~\\

\begin{figure}[ht] 

  \begin{minipage}[l]{.48\textwidth} 
   \begin{center}
   \includegraphics[width=\textwidth]{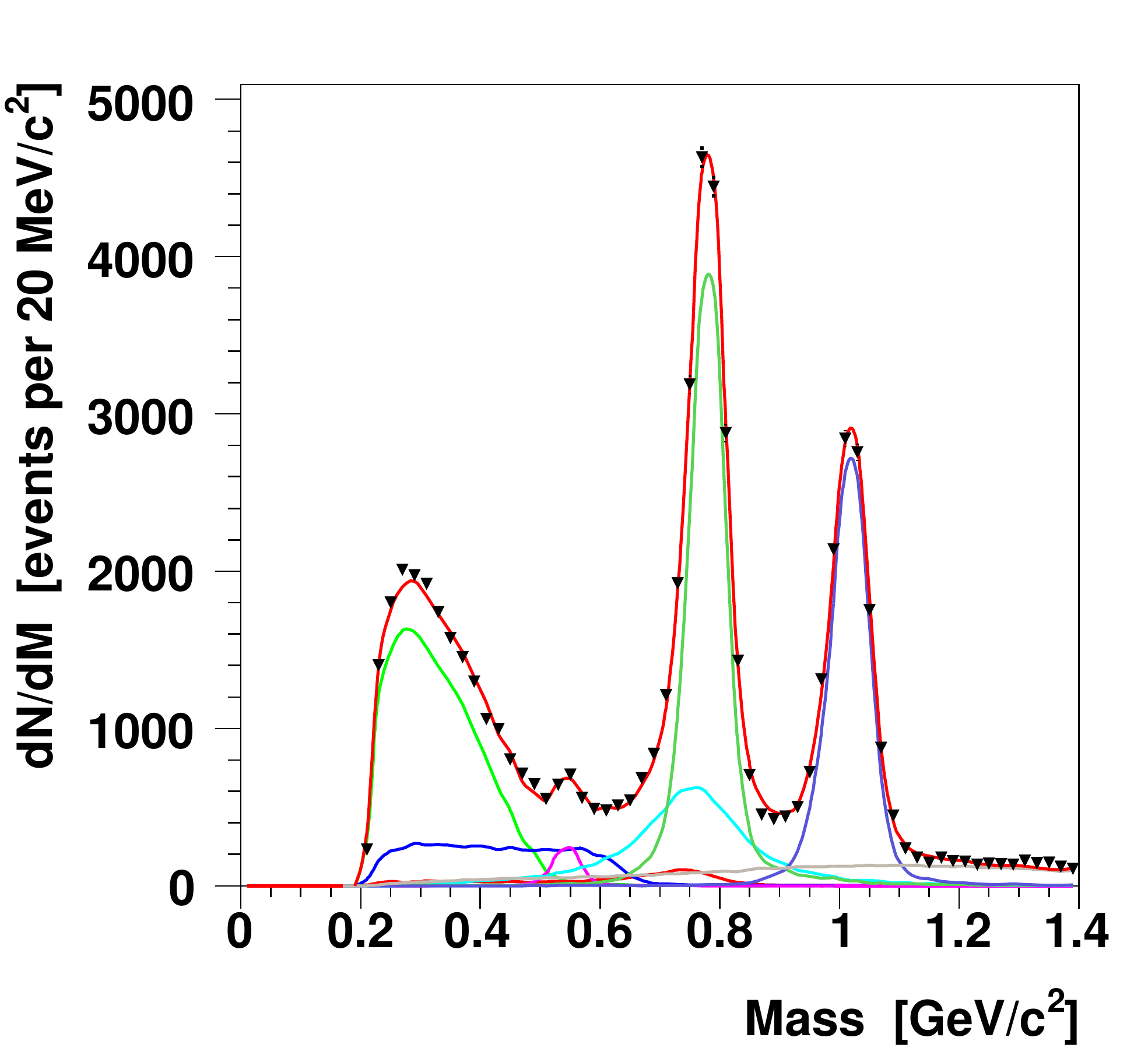} 
   \end{center} 
\vspace{-0.6cm}
\caption{\label{fig:massSpectrum} Fit on the target-integrated mass spectrum.}
\vspace{1.25cm}
  \end{minipage} 
  \hspace{0.03\textwidth}
   \begin{minipage}[r]{.45\textwidth} 
    \begin{center} 
    \includegraphics[width=\textwidth]{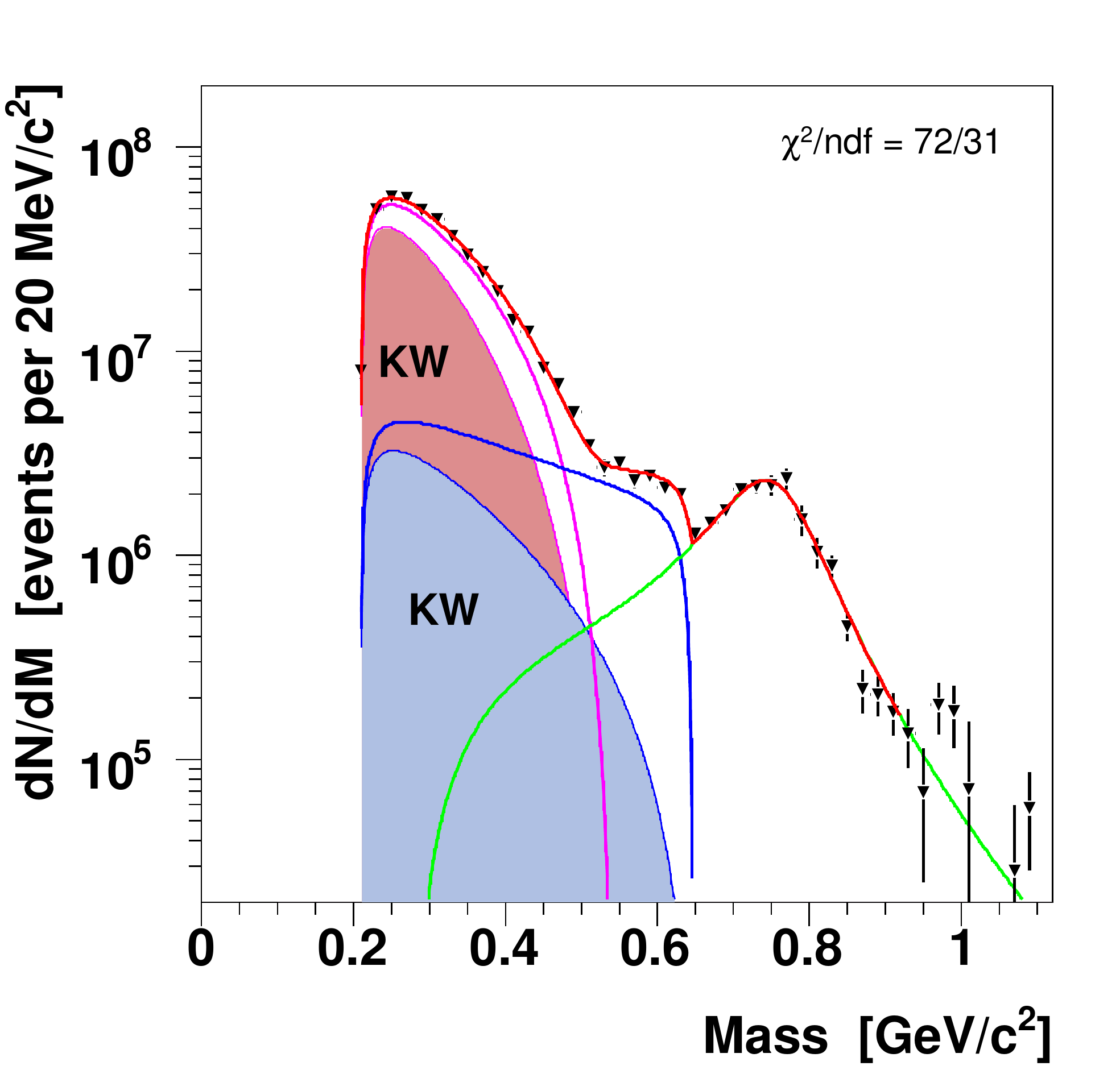} 
    \end{center} 
\vspace{-0.6cm}
\caption{\label{fig:formFactors} Fit on the acceptance-corrected mass spectrum. The shaded areas indicate the Kroll-Wada (KW) expectations for the $\eta$ and $\omega$ Dalitz decays for point-like particles, defined by QED~\cite{KW}.}
  \end{minipage} 

\end{figure}

\noindent The preliminary results presented here are obtained from the analysis of the data collected during three days within the 2004 proton run, with a system of nine sub-targets of different nuclear species, simultaneously exposed to an incident 400~GeV proton beam. 
The first step of the analysis is a fit of the reconstructed mass spectrum (see \figurename~\ref{fig:massSpectrum}). The fit is performed comparing the data with the superposition of the expected sources (estimated via the MC). These sources are the Dalitz decays of the $\eta$, $\omega$ and $\eta'$ mesons, the 2-body decays of the $\eta$, $\rho$, $\omega$ and $\phi$ and the contribution of the open charm process. Before the comparison between data and MC is performed, the small component of combinatorial background (originating from $\pi$ and $K$ decays) is subtracted from the real data: its shape is estimated with an event mixing technique, while its normalization is established fixing the LS (like-sign) component coming out from the mixing, to the LS component of the data (containing no signal from correlated pairs at the SPS energies). The background accounts for less than $10\%$ of the integrated mass spectrum below 1.4~GeV$/c^2$. The comparison between mixed and real sample gives an average uncertainty of $10\%$, for both the $(++)$ and the $(--)$ components: given the rather low level of the combinatorial background contribution, this uncertainty hardly affects the evaluation of the shape for the Dalitz decays under study.

The processes which are not of interest for the measurement of the form factors of the $\eta$ and $\omega$ are then subtracted from the real data, using the normalizations fixed by the fit. Only the Dalitz decays of $\eta$ and $\omega$ are retained, together with the two-body decay of the $\rho$: this latter is left in, making it much easier to control the systematics due to the small contribution coming from the low-mass tail of the $\rho$ in the mass region of interest here.  The mass spectrum resulting after the subtraction is then corrected for the effects of geometrical acceptance and reconstruction efficiency. In order to do this, a correction profile as a function of the mass is built, weighting the profiles obtained from the MC for the three processes separately. After the correction for the acceptance~$\times$~efficiency effects, the resulting mass spectrum $-$ now free from any acceptance distortion $-$ is compared with the superposition of the expected line shapes of the three processes $\eta \to \mu^+\mu^-\gamma~$, $\omega \to \mu^+\mu^-\pi^0~$ and $\rho \to \mu^+ \mu^-$.

~\\
\noindent \figurename~\ref{fig:formFactors} shows the fit on the acceptance-corrected mass spectrum (black triangles), having isolated the contribution of the three processes $\eta \to \mu^+\mu^-\gamma~$, $\omega \to \mu^+\mu^-\pi^0~$ and $\rho \to \mu^+ \mu^-$. The fit function is a superposition of the expected line shapes for the three processes, described by the following functions:

\begin{footnotesize}
\vspace{-0.2cm}
\begin{eqnarray}
 \left. \frac{\mathrm{d}N}{\mathrm{d}M} \right|_{\eta_\mathrm{Dal}} & \propto & 2M \times \frac{2}{3} \frac{\alpha}{\pi} \frac{\Gamma(\eta \rightarrow
\gamma \gamma)}{M^2} \times \left( 1 - \frac{M^2}{m_\eta^2} \right)^3 \left( 1 +
\frac{2m_\eta^2}{M^2} \right) \left( 1 - \frac{4m_\eta^2}{M^2} \right)^{1/2}
\times \left| F_\eta(M^2) \right|^2 ~~~~~~
\end{eqnarray}
\end{footnotesize}

\begin{footnotesize}
\vspace{-0.5cm}
\begin{eqnarray}
 \left. \frac{\mathrm{d}N}{\mathrm{d}M} \right|_{\omega_\mathrm{Dal}} & \propto &
 2M \times \frac{\alpha}{3\pi} 
 \frac{\Gamma(\omega \rightarrow \gamma \pi^0)}{M^2} \times \left( 1 +
\frac{2m_\mu^2}{M^2} \right) \times \nonumber \\
 &  & \times \left( 1 - \frac{4m_\eta^2}{M^2} \right)^{1/2} \times
 \left[ \left( 1 + \frac{M^2}{m_\omega^2 -m_{\pi^0}^2} \right)^2 -
\frac{4m_\omega^2 M^2}{(m_\omega^2 -m_{\pi^0}^2)^2} \right] \times
 \left| F_\omega(M^2) \right|^2~~~~~~~~~~~~~~
\end{eqnarray}
\end{footnotesize}

\begin{footnotesize}
\vspace{-0.5cm}
\begin{eqnarray}
 \left. \frac{\mathrm{d}N}{\mathrm{d}M} \right|_\rho & \propto &
 \frac{\alpha^2 m_\rho^4}{3(2\pi)^4} \times
 \left( 1 - \frac{4m_\pi^2}{M^2} \right)^{3/2} \times \nonumber \\
 & & \times \left( 1 - \frac{4m_\mu^2}{M^2} \right)^{1/2} 
 \left( 1 + \frac{2m_\mu^2}{M^2} \right)
 \times (2\pi MT)^{3/2} \times
 \frac{1}{(M^2-m_\rho^2)^2 + (M \Gamma_\mathrm{tot}^0)^2}~~~~~~~~~~~~~~~~
\end{eqnarray}
\vspace{0.3cm}
\end{footnotesize}

\begin{figure}[ht] 
\vspace{-0.3cm}
  \begin{minipage}[l]{.48\textwidth} 
   \begin{center}
   \includegraphics[width=\textwidth]{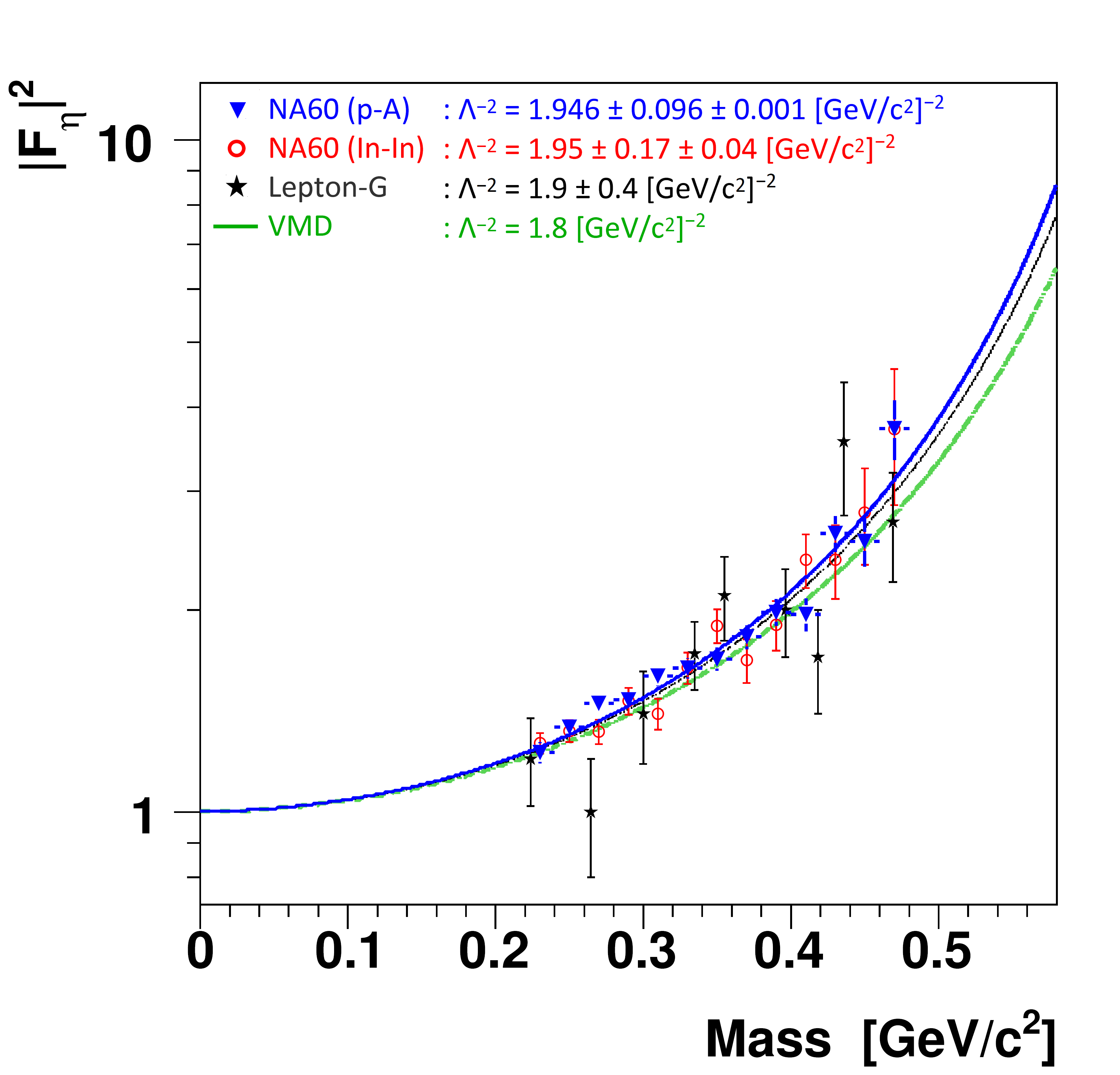} 
   \end{center} 
  \end{minipage} 
  \hspace{0.03\textwidth}
   \begin{minipage}[r]{.48\textwidth} 
    \begin{center} 
    \includegraphics[width=\textwidth]{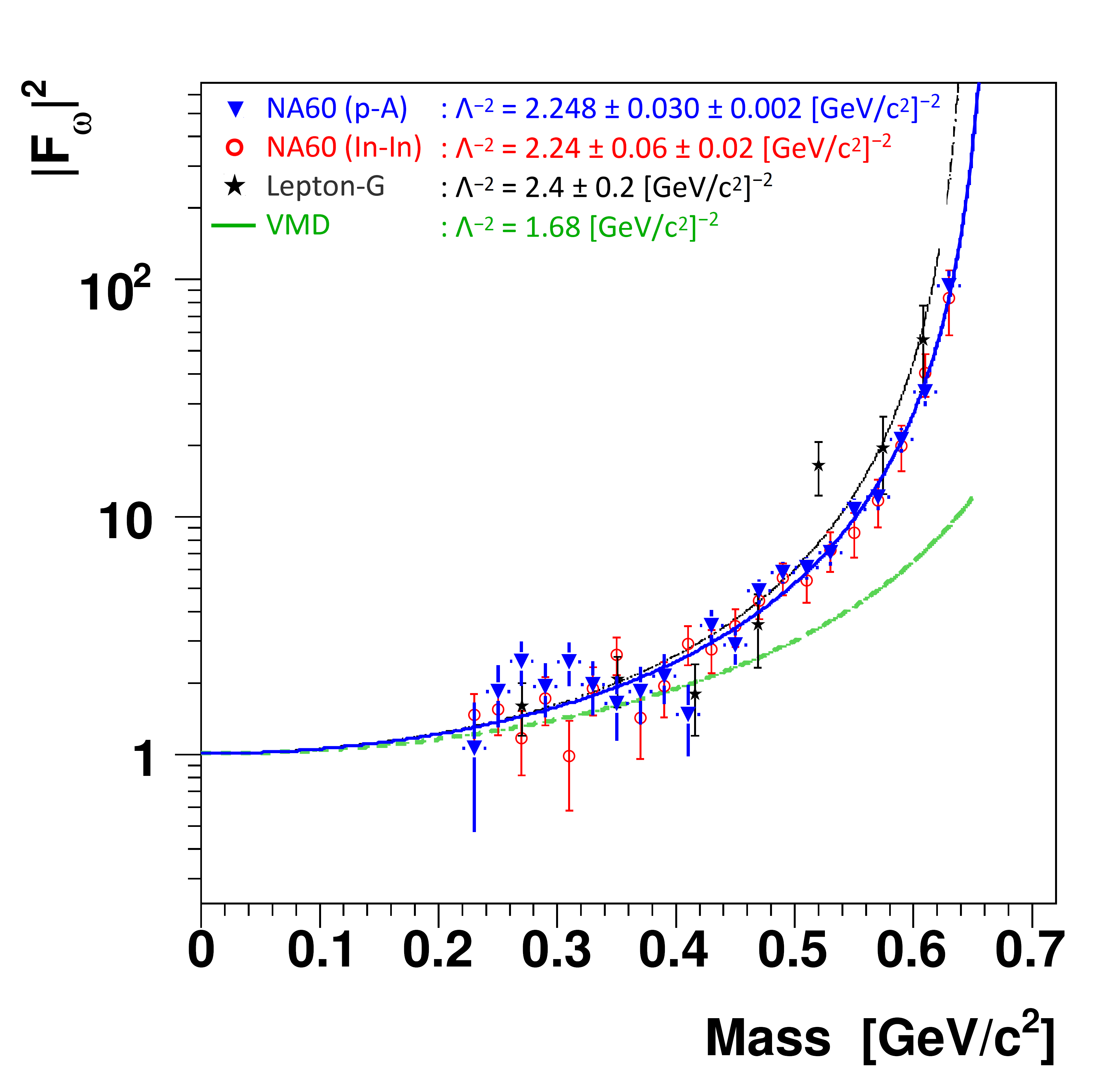} 
    \end{center} 
  \end{minipage} 
  \vspace{-0.3cm}
  \caption{Electromagnetic transition form factors for the $\eta$ (left) and $\omega$ (right) mesons as a function of $M$.}
\label{fig:formFactorsDisentagled}
  \vspace{-0.3cm}
\end{figure}

\noindent In the fit, the following parameters are left free:
\begin{itemize}
 \item the three normalizations, one for each of the line shapes above
 \item the parameters $\Lambda_\eta^{-2}$ and $\Lambda_\omega^{-2}$, contained
 in the form factors $\left| F_\eta(M^2) \right|^2$ and $\left|
F_\omega(M^2) \right|^2$. 
\end{itemize}
Concerning the line shape for the $\rho \to \mu^+ \mu^-$ decay, the parameters
$\Gamma_\rho$ (entering in the definition of $\Gamma_\mathrm{tot}$), $m_\rho$ and
the effective temperature $T_\rho$ are fixed respectively to $\Gamma_\rho =
151$~MeV (from the PDG), $m_\rho = 768$~MeV/$c^2$ (also from the PDG) and
$T_\rho = 170$~MeV (consistent with the expected value obtained by statistical model fits on particle ratios in $pp$ interactions).
All the parameters involved in the analysis procedure have been varied to study their influence on the fits and determine the systematic uncertainties on the results: the weights defining the acceptance$\times$efficiency correction, the fit range (and the choice of the free parameters), the value for the $\sigma_{\eta'}/\sigma_{\omega}$ ratio, the charge correlation parameter for the combinatorial background and finally the level of the open charm contribution. The analysis leads to the following values:
\begin{eqnarray}
 \Lambda_\eta^{-2} = 1.946 \pm 0.096~\mathrm{(stat.)}~\pm 0.015~\mathrm{(syst.)}~(\mathrm{GeV}/c^{2})^{-2} \nonumber \\
 \Lambda_\omega^{-2} = 2.248 \pm 0.030~\mathrm{(stat.)}~\pm 0.009~\mathrm{(syst.)}~(\mathrm{GeV}/c^{2})^{-2} \nonumber
\end{eqnarray}
Once the final fit parameters and their errors are fixed, the contributions of the $\eta \to \mu^+\mu^-\gamma~$ and $\omega \to \mu^+\mu^-\pi^0~$ processes are disentangled, making it then possible to present the two form factors in the usual way, in order to compare them with the previous experimental points, see \figurename~\ref{fig:formFactorsDisentagled}. The pole parameters and their errors as obtained from the combined fit to both Dalitz decays are shown as inserts. Both figures include the NA60 data obtained in peripheral~In-In, the Lepton-G data, and the expectations from VMD for comparison. Within the errors, perfect agreement between the three data sets is seen in both cases. Irrespective of the much reduced errors, the form factor of the $\eta$ is still close to the expectation from VMD. The form factor of the $\omega$, on the other hand, strongly deviates from the VMD, showing a relative increase close to the
kinematic cut-off by a factor of $\sim10$.

~\\
\noindent In conclusion, we have presented a new measurement of the electromagnetic transition form factors of the $\eta$ and $\omega$ mesons. The analysis has been performed by studying the Dalitz decays $\eta \to \mu^+\mu^-\gamma$ and $\omega \to \mu^+\mu^-\pi^0$ in p-A collisions at 400~GeV/$c$ with the NA60 apparatus. The new preliminary values presented here are in good agreement with those already available; improving the precision of the measure by a factor $\sim 2$, they confirm on more solid grounds the discrepancy between the VMD prediction for the $\omega$ and the experimental observations.

\end{document}